\documentclass[aps,prd,reprint,superscriptaddress,nofootinbib,amsfonts,amssymb,amsmath,twocolumn]{revtex4-1}

\usepackage[utf8x]{inputenc}
\usepackage[dvipdf,dvips,dvipdfmx]{graphicx}
\usepackage{float}
\usepackage{wrapfig}
\usepackage{bm}
\usepackage{color}
\usepackage{comment}
\usepackage{xcolor}
\usepackage[normalem]{ulem}
\usepackage{hyperref}
\hypersetup{
colorlinks=true,
citecolor=blue,
citebordercolor=red,
linktoc=all,
linkcolor=blue,
urlcolor=blue
}
\usepackage[capitalise]{cleveref}
\usepackage{orcidlink}

\catcode`\@=11
\def\lsim{\mathrel{\mathpalette\@versim<}}
\def\gsim{\mathrel{\mathpalette\@versim>}}
\def\@versim#1#2{\vcenter{\offinterlineskip
\ialign{$\m@th#1\hfil##\hfil$\crcr#2\crcr\sim\crcr } }}
\catcode`\@=12

\newcommand{\be}{\begin{equation}}
\newcommand{\ee}{\end{equation}}
\newcommand{\bea}{\begin{eqnarray}}
\newcommand{\eea}{\end{eqnarray}}

\crefformat{equation}{Eq.~(#2#1#3)}
\crefformat{table}{Tab.~(#2#1#3)}
\crefformat{figure}{Fig.~(#2#1#3)}
\crefformat{appendix}{App.~(#2#1#3)}
\crefformat{section}{Sec.~(#2#1#3)}
\crefformat{subsection}{Subsec.~(#2#1#3)}

\begin{document}

\title{On the Renormalization Group flow of distributions}
\author{Astrid Eichhorn\,\orcidlink{0000-0003-4458-1495}}
\email{eichhorn@thphys.uni-heidelberg.de}
\affiliation{Institute for Theoretical Physics, Heidelberg University, Philosophenweg 16, 69120 Heidelberg, Germany}
\author{Aaron Held\,\orcidlink{0000-0003-2701-9361}}
\email{aaron.held@phys.ens.fr}
\affiliation{
Institut de Physique Théorique Philippe Meyer, Laboratoire de Physique de l’\'Ecole normale sup\'erieure (ENS), Universit\'e PSL, CNRS, Sorbonne Universit\'e, Universit\'e Paris Cité, F-75005 Paris, France
}

\begin{abstract}
Renormalization Group flows relate the values of couplings at different scales. Here, we go beyond the Renormalization Group flow of individual trajectories and derive an evolution equation for a distribution on the space of couplings. This shift in perspective can provide new insights, even in theories for which the Renormalization Group flow of individual couplings is well understood.
\\
As a first application, we propagate errors under the Renormalization Group flow. Characteristic properties of an error distribution, such as its 
maximum or highest density region, cannot be propagated at the level of individual couplings, but require our evolution equation for the distribution on the space of couplings. We demonstrate this by calculating the most probable value for the metastability scale in the Higgs sector of the Standard Model.
\\
Our second application is the emergence of structure in sets of couplings. We discover that infrared-attractive fixed points do not necessarily attract the maximum of the distribution when the Renormalization Group is evolved over a finite range of scales. Instead, sharply peaked maxima can build up at coupling values that cannot be inferred from individual trajectories. We demonstrate this emergence of structure for the Standard Model, starting from a broad  distribution at the Planck scale. The Renormalization Group flow favors the phenomenological ordering of third-generation Yukawa couplings and, strikingly, we find that the most probable values for the Abelian hypercharge and Higgs quartic coupling lie close to their observed values.
\end{abstract}
\maketitle
\noindent\emph{Introduction}:
In quantum field theory (QFT), the Renormalization Group (RG) flow determines the evolution of couplings from one energy scale to another. It is standard to consider a single initial condition for each coupling and calculate the corresponding individual RG trajectory. 
Here, we develop a different perspective.
Instead of single trajectories, we consider distributions on the space of couplings. 
The RG evolution of distributions is important whenever initial conditions are not known exactly.
Important applications include experimental as well as theoretical uncertainties.
\\

\noindent\emph{Evolution equation for distributions
in coupling space}:
Given 
the beta functions $\{\beta_{g_i}= \partial_t g_i(t)\}$ with $t= \ln(k/k_0)$ that determine the RG evolution of individual couplings $g_i$, we derive the partial differential equation governing the RG evolution of the distribution $P(\vec{g},t)$ in the supplementary material and obtain
\bea
\partial_t P(\vec{g},t)&=& -\sum_i \left(\frac{\partial \beta_{g_i}}{\partial g_i}P+ \beta_{g_i} \frac{\partial}{\partial g_i} P\right)
.\label{eq:partialtP}
\eea

Eq.~\eqref{eq:partialtP} can be solved by the method of characteristics. First, we solve $\beta_{g_i} = \partial_t g_i(t)$ 
with the vector of initial conditions $\vec{g}(t=t_0)=\vec{g}_0$ to obtain integrated RG trajectories $\vec{g}(t, \vec{g}_0)$. Second, we invert the relation $\vec{g}(t, \vec{g}_0)$ 
to obtain $\vec{g}_{0}(t,\vec{g})$, i.e., we express the initial condition in terms of the values of the couplings obtained after evolving from the initial RG time $t_0$ to the final RG time $t$.
Together with an initial distribution $P(\vec{g}(t_0),t_0)=P_0(\vec{g}_0)$
we arrive at
\begin{eqnarray}
	P(\vec{g},\,t) = 
	P_0\left(\vec{g}_0(t,\vec{g})\right)\cdot e^{\left[
		 - \int_{t_0}^{t} dt'\,{
		 	\sum_i \frac{\partial \beta_{g_i}}{\partial g_i}  
		 }
	\right]}
	\label{eq:formal-solution}
	\;.
\end{eqnarray}
This general solution 
consists of two factors. The first is the initial distribution $P_0$, depending on the final values $\vec{g}(t)$ through $\vec{g}_0(t, \vec{g})$.   
This factor accounts for the individual RG trajectories of couplings. It contains non-linearities -- e.g., a Gaussian distribution in $\vec{g}_0$ does not in general correspond to a Gaussian distribution in 
$\vec{g}$ -- but it maps the maximum of the distribution at one scale to the maximum at another scale through one individual RG trajectory. 
For instance, for a one-dimensional example, we have that $0 = \partial_g P(g_0(t, g)) = \left(\partial_{g_0} P(g_0(t,g))\right)\partial_g g_0(t,g)$, which is solved by $\partial_{g_0} P(g_0)=0$. 
The second factor, the exponential of the integrated derivative of the beta function, introduces additional nonlinearities into the RG evolution of $P(g,t)$. It is crucial for the maxima of the evolved distribution, because 
\begin{align}
0= \partial_g P(g,t) 
&= 
\left(\partial_{g_0}P(g_0)\right) \partial_g g_0(t,g)\cdot 
e^{\left[ 
    - \int_{t_0}^{t} dt'\,{
        \partial_g \beta_g
    }
\right]}
\nonumber\\
&\quad
+ P_0(g_0(t,g))\cdot 
\partial_g \, 
e^{\left[ 
    - \int_{t_0}^{t} dt'\,{
        \partial_g \beta_g
    }
\right]}
\;.
\label{eq:maximumflow}
\end{align}
Requiring $\partial_{g_0}P(g_0)=0$ does not solve this equation. Thus, in general, the maximum of the distribution at some RG scale $t$ is \emph{not} the value obtained from the individual RG trajectory starting at the maximum $g(t_0)= g_{\rm max}$ at the initial RG scale.
\\

\noindent\emph{Application 1: Error propagation under the RG flow}: As a first consequence of Eq.~\eqref{eq:formal-solution}, proper error propagation under the RG flow cannot be achieved by RG evolving individual values of couplings.\\
The values of couplings may be uncertain at a given scale, e.g., because they are inferred from experiment, or because the values are calculated from a UV theory to finite order in an approximation. 
To propagate these errors from one scale to another, one must propagate the distribution that characterizes the uncertainty at the initial scale. Due to Eq.~\eqref{eq:maximumflow}, one obtains incorrect results, if one calculates, e.g., the central value at one scale and then RG evolves the coupling using its beta function. 

\begin{figure}
\includegraphics[width=\linewidth]{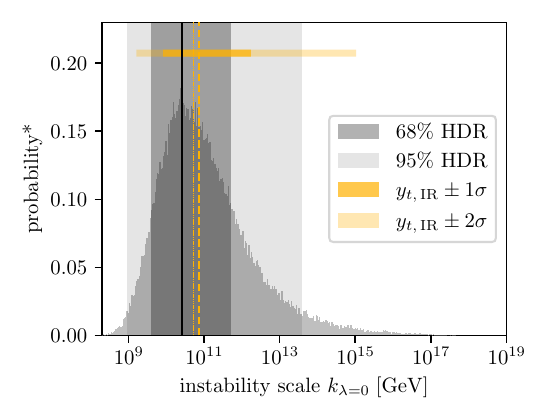}
\caption{\label{Fig:higgs-stability}
    Comparison of likelihood estimates for the instability scale $k|_{\lambda=0}$ due to the experimental uncertainty on the top quark mass $M_t$ obtained at three-loop order. The most likely value (continuous black vertical line) lies below the RG evolved central value (dashed yellow vertical line). The shift between the two values is significantly larger than the shift produced by including three-loop terms in the top-Yukawa running (compare for thin dashed yellow vertical line indicating the RG evolved central value without three-loop contributions to the running of $y_t$). \\
    Similarly, the boundaries of the 68(95)\% highest density regions (gray vertical bands) are distinct from the RG evolved 1(2) $\sigma$ regions.
}
\end{figure}

We exemplify this on the Higgs sector of the SM. 
Starting from the central value of the SM couplings inferred from experiments and RG evolving towards the UV,
the Higgs quartic coupling $\lambda$ crosses zero at a scale $k|_{\lambda=0}\approx 10^{11}\, \rm GeV$~\cite{Bezrukov:2012sa,Degrassi:2012ry}. A negative quartic coupling is interpreted as a signal of an unstable or metastable Higgs potential~\cite{Hung:1979dn,Politzer:1978ic,Lindner:1988ww,Sher:1988mj,Arnold:1989cb,Ford:1992mv,Sher:1993mf,Altarelli:1994rb,Casas:1994qy,Espinosa:1995se,Isidori:2001bm,Ellis:2009tp,Degrassi:2012ry, Bezrukov:2012sa, Buttazzo:2013uya}. Solutions invoking new physics have been proposed, see, e.g.,~\cite{Lebedev:2012zw,Elias-Miro:2012eoi,Chao:2012mx,Gabrielli:2013hma,Gies:2014xha,Eichhorn:2015kea,Ballesteros:2016xej} and references therein. One may think of $k|_{\lambda=0}$ as an estimate for the mass-scale of new physics.
However, the experimental values have uncertainties, which are relatively large for the top Yukawa coupling. To determine  bounds on the Higgs and the top mass that avoid meta-stability, it is of course correct to evolve individual RG trajectories, as is being commonly done~\cite{Elias-Miro:2011sqh,Bezrukov:2012sa,Degrassi:2012ry,Buttazzo:2013uya, Hiller:2024zjp}.
However, to determine the most likely value of $k|_{\lambda=0}$, we must evolve the full distribution. We demonstrate that the two are different in Fig.~\ref{Fig:higgs-stability}, where we use three-loop\footnote{To be precise, we use three-loop running for the couplings $(g_Y,\,g_2,\,g_3,\,y_t)$ and two-loop running for $(y_b,\,y_\tau,\,\lambda)$. The three-loop contributions in $(y_b,\,y_\tau,\,\lambda)$ are negligible, just like all contributions of any of the other SM couplings, see~\cite{Bezrukov:2012sa}.} running as in~\cite{Bezrukov:2012sa} for the SM couplings
\begin{align}
	g = (
		\underbrace{g_Y,\,g_2,\,g_3}_\text{gauge},\,
		\underbrace{y_t,\,y_b,\,y_\tau
        }_\text{Yukawa},\,
		\underbrace{\lambda}_\text{Higgs}
	)\;.\label{eq:couplinglist}
\end{align}
We use the matching conditions of \cite{Bezrukov:2012sa} to obtain initial values for the RG evolution at a common matching scale $k=M_{t,\mathrm{central}}$. We take the central values of the couplings from \cite{ParticleDataGroup:2024cfk}. We take into account the experimental uncertainty on the top mass, which we interpret as a Gaussian distribution for $y_t(M_{t,\mathrm{central}})$ at the matching scale. For simplicity, we neglect all other experimental uncertainties.

We find that the most probable value of $k|_{\lambda=0}$ is $2.6\times10^{10}\,\text{GeV}$, which is different from the RG evolved central value by a factor of three. The mean value of $k|_{\lambda=0}$ also differs from both other values substantially. In addition, the 1(2)-$\sigma$ intervals of the initial Gaussian distributions are mapped to intervals containing 68 \% (95 \% etc.) of the couplings values. However, these intervals carry no special meaning in the UV distribution. In fact, they are distinct from the 68 \% (95 \% etc.) highest-density regions, which actually characterize the distribution. 
The roughly 1.3 \% of trajectories that reach the Planck scale without developing a negative quartic coupling are, of course, correctly identified from studies following individual trajectories \cite{Elias-Miro:2011sqh,Bezrukov:2012sa,Degrassi:2012ry,Buttazzo:2013uya,Hiller:2024zjp}.

Thus, while some conclusions can be drawn from RG evolving individual trajectories, not all relevant questions can be answered in this way. For instance, one can calculate the value of $k|_{\lambda=0}$ associated to the central value of the experimental results from an individual RG trajectory. However, one cannot answer what the most probable value of $k|_{\lambda=0}$ is. The latter value, when interpreted as a candidate for a scale of new physics, may be relevant for many theoretical or even experimental endeavors.

For general cases, we glean from our example that, i) the maximum of the distribution is not equal to the RG evolved central value, ii) the highest-density regions differ from the RG evolved $n-\sigma$ boundary values, iii) the mean of the distribution is not equal to the RG evolved mean value, iv) mean value and maximal value are in general distinct, because the evolved distribution is in general asymmetric.
\\

\begin{figure*}[!t]
\begin{centering}
\includegraphics[width=\linewidth]{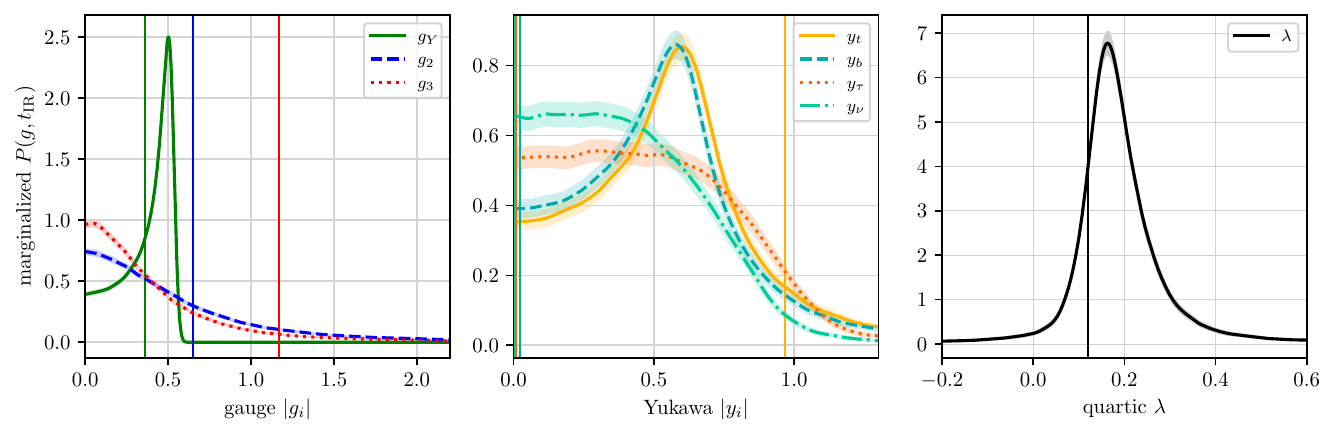}
\end{centering}
\caption{\label{fig:SMnumericalflow}
We show the IR distribution resulting from a unit-width isotropic Gaussian distribution at the Planck scale (see supplementary material for the corresponding plots of the UV distribution) marginalized to its dependence on individual couplings. 
For comparison, we also highlight the experimental SM values (see vertical lines in the lower panel). The shaded bands indicate statistical error estimates due to the finite number of samples approximating the distributions.
}
\end{figure*}

\noindent\emph{Application 2: Structures from broad UV distributions}:
We now turn our attention to RG flows from the UV to the IR. We explore the distribution and its maxima for the SM (see~\cref{eq:couplinglist}), focusing on third-generation Yukawa couplings only and including a right-handed neutrino with its Yukawa coupling $y_{\nu_{\tau}}$. Restricting to one-loop RG flows, we investigate 
the emergence of structure.
While the formal solution in Eq.~\eqref{eq:formal-solution} is fully general, obtaining an explicit solution still requires us to solve the set of $\beta$ functions which govern individual RG trajectories. For multidimensional theory spaces, this requires numerical solution of the respective set of ordinary differential equations.
We solve this numerical problem by Monte-Carlo sampling the initial UV distribution and then individually evolving the resulting large set of RG trajectories, see~\cref{fig:SMnumericalflow}.

We choose a multidimensional, isotropic Gaussian distribution with unit width, centered about the origin, as the initial distribution in the UV.
The RG flow towards the IR features Landau poles in the one-loop beta functions for the non-Abelian gauge couplings, which signal the onset of nonperturbative physics. We remove all RG trajectories that end in Landau poles during the RG flow from the Planck scale to the electroweak scale. This removes large absolute values of $g_2$ and $g_3$ (and corresponding large values of $\Lambda_{\rm QCD}$ as well as non-perturbative $SU(2)$ effects) from the initial distributions at the Planck scale; and results in a slightly asymmetric distribution for $\lambda$ at the Planck scale, see supplementary material.

We aim to discover whether the RG flow of the distribution generates nontrivial structure \emph{which is not put in by hand at the UV scale.}
Such structures, e.g., nontrivial maxima, can in principle serve as probabilistic explanations of measured values, if nontrivial maxima lie close to measured values for generic choices of the UV distribution. What may be considered as a generic choice to some extent  is subjective. However, we consider an isotropic Gaussian distribution of unit width and centered about the origin generic, as opposed to, e.g., Gaussian distributions that are sharply peaked.

We find a nontrivial distribution with a clearly developed peak for the Abelian gauge coupling at $g_Y \approx 0.5$ with the 68 \% highest-density interval being $[0.33,0.54]$ (compared to the experimental result $g_Y =0.358$)\footnote{A better match with experimental results can be obtained by slightly adjusting the width (or shape) of the initial distribution.}.
Given that $g_Y=0$ is an IR-attractive fixed point and also the maximum of the UV distribution, one may have expected that $g_Y=0$ 
stays the most probable value at all scales. That this is not the case is a consequence of the nonlinearities in Eq.~\eqref{eq:formal-solution}.

\begin{figure}
\centering
\includegraphics[width=\linewidth]{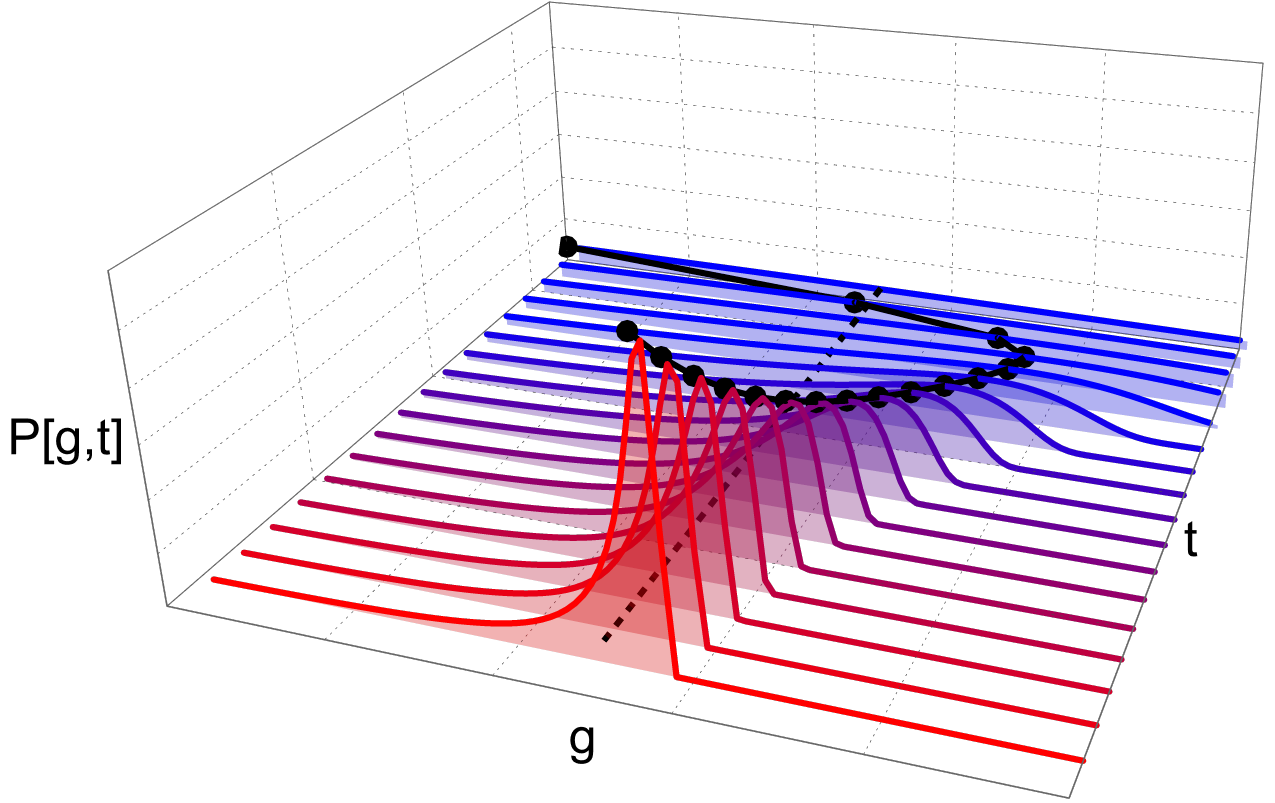}
\hfill
\caption{RG flow of probability distributions from the UV (blue and uppermost line) to the IR (red and lowermost line), for an example with $\beta_g = -g + \beta_1\, g^2$, with $\beta_1>0$ and thus an IR-attractive fixed point. Any given RG trajectory for the coupling is attracted towards $g_{\ast} = 1/\beta_1$, which is an IR attractive fixed point. Nevertheless, the same is not true for the maximum of the distribution. 
Due to the non-linearities in Eq.~\eqref{eq:formal-solution}, 
the maximum (indicated by black dots), which is initially chosen to be smaller than $g_{\ast}$, first evolves to values larger than $g_{\ast}$ and finally approaches $g_{\ast}$ from above.
This exemplifies that IR attractive fixed points \emph{do not attract the maximum of the distribution monotonically.} 
\label{fig:evolution-of-distributions_IRFP}
}
\end{figure}

At one-loop order, the RG evolution of the Abelian gauge coupling $g_Y$ decouples from the other SM couplings and we can thus analytically solve the RG evolution of its distribution. We work with $\alpha = \frac{g_Y^2}{4\pi}$ and consider
\begin{align}
	\beta_{\alpha} = 
    \beta_1\,\alpha^2\;,
    \label{eq:beta-1D}
\end{align}
where $\beta_1 = \frac{1}{4\pi}\frac{41}{3}$ is the one-loop  coefficient.
This beta function can be solved with initial condition $\alpha(t_0)=\alpha_0$ to obtain the well-known RG trajectories
\be
	\alpha(t, \alpha_0) = \frac{\alpha_0}{1-\alpha_0\, \beta_1\left( t-t_0\right)}
    \;.
\ee
We invert this to express $\alpha_0$ in terms of $\alpha$
\be
\alpha_0(t, \alpha) = 
\frac{\alpha}{1+\alpha\, \beta_1 \left(t-t_0\right)}.
\label{eq:initialcond}
\ee
For the beta-function Eq.~\eqref{eq:beta-1D} and the relation \eqref{eq:initialcond} between the initial condition $\alpha_0$ and the final value $\alpha(t)$, $P(\alpha(t),t)$ in Eq.~\eqref{eq:formal-solution} becomes
\be
	P(\alpha(t),t) = 
	P_0\left(\alpha_0(t,\alpha)\right) \cdot
	\left(\frac{\alpha_0(t,\alpha)}{\alpha}\right)^2
	\;.
    \label{eq:formal-solution-1D}
\ee
For the example $P_0 = \frac{1}{\sqrt{2\pi}\sigma}e^{-\frac{\alpha^2}{2 \sigma^2}}$, the maximum lies at
\begin{align}
\alpha_{\rm max} &= \frac{1+4(t_0-t)^2\beta_1^2\, \sigma^2 - \sqrt{1+ 8(t_0-t)^2\beta_1^2\, \sigma^2}}{4(t_0-t)^3 \beta_1^3\, \sigma^2}
\;.
\end{align}
The scale evolution of $\alpha_{\rm max}$ is clearly non-monotonic, and only approaches $\alpha_{\rm max}=0$ in the limit $(t_0-t) \rightarrow \infty$. A similar behavior occurs even if the IR attractive fixed point lies at a nonzero coupling, see Fig.~\ref{fig:evolution-of-distributions_IRFP}.

In the right panel of Fig.~\ref{fig:SMnumericalflow}, one sees that, similarly to the Abelian hypercharge, the Higgs quartic coupling exhibits a relatively sharp peak not far from the experimental value. Here, the explanation is more intricate, because $\beta_{\lambda}$ depends on other SM couplings. In particular, there is an IR attractive quasi-fixed point in $\beta_{\lambda}$, i.e., if all other couplings are held fixed, $16\cdot\lambda =3g_2^2 + g_Y^2 - 4y_t^2 - \sqrt{80 y_t^4-8 g_Y^2\, y_t^2 - 24 g_2^2 \, y_t^2 - 15 g_Y^4-2g_2^2\, g_Y^2 - 3g_2^2}$ is an IR attractive point. Because the other couplings depend on the scale, this is not a true fixed point, instead, it constitutes a scale-dependent attractor for the RG flow of the Higgs quartic coupling.
For the IR values of $y_t$, $g_Y$ and $g_2$, the fixed point lies at $\lambda \approx 0.3$. Thus, the sharp peak in $\lambda$ is due to a non-monotonic (see above) approach to this quasi-fixed point, similar to the behavior demonstrated in~\cref{fig:evolution-of-distributions_IRFP}.

In the central panel of~\cref{fig:SMnumericalflow}, we see that, in the Yukawa sector, structure emerges that matches the SM ordering  $y_t> y_b > y_{\tau} >y_{\nu}$, albeit only qualitatively, not quantitatively\footnote{Here we assume a mass-generation for neutrinos through a tiny Yukawa coupling; not a see-saw mechanism.}.
First, the distributions for top- and bottom Yukawa coupling develop clear maxima at non-zero values. Second, the maximum for the top Yukawa lies at larger values than that for the bottom Yukawa, although the effect is quantitatively very small. Third, the Yukawa couplings of the tau and of the neutrino develop approximately flat plateaus, so they are more likely to be close to zero than top and bottom Yukawa.  This occurs because only quarks couple to the SU(3) gauge couplings, which makes the flow more nonlinear for the quark Yukawas; in contrast the flow for the lepton Yukawas is slow and only results in a mild slimming of the distribution, see supplementary material for a quantitative demonstration.
Fourth, the distribution for the tau is broader than that for the neutrino, such that the neutrino Yukawa is most likely among all four Yukawas to be tiny.
Thus, there is a certain degree of universality in SM-like realizations of the couplings, e.g., it is more likely for quark and lepton masses to differ rather than be the same, even if we choose the same broad UV distribution for all four Yukawa couplings.
\\

\noindent\emph{Conclusions and outlook}:
In this paper, we have shifted the perspective from the RG flow of individual couplings to RG flows of distributions of couplings. We have discovered that this shift in perspective can provide novel insights, even for theories for which the RG flow of individual couplings is extremely well-investigated, e.g., for the SM. 

As an example, we have discovered non-trivial structure in the SM couplings, with sharp peaks in the Abelian hypercharge and Higgs quartic coupling not far from their experimental values; and reproduced the qualitative ordering of Yukawa couplings in the third generation. 
Our results thus provide a new perspective on the long-standing problem of explaining the values of couplings in the SM. While not providing sharp predictions, our probabilistic point of view shows that nontrivial, SM-like structures  are likely.

We also expect that further developing our new perspective could be fruitful  for quantum theories of gravity, in particular for such theories that, like string theory, do not have a unique realization. A ``scan" over different compactifications gives rise to a distribution of coupling values \cite{Cvetic:2022fnv,Marchesano:2024gul,Constantin:2024yxh,Constantin:2024yaz} . Based on a different mechanism, asymptotically safe gravity gives rise to an allowed interval for Yukawa couplings at the Planck scale \cite{Eichhorn:2017ylw,Eichhorn:2018whv}, i.e., a distribution that is limited from above, see also \cite{Eichhorn:2022gku}.
Our results show that qualitatively (albeit not quantitatively) SM-like structures emerge with high probability from reasonably generic distributions at the Planck scale. 

At the heart of these results lies the fact that, while individual RG trajectories are always attracted towards infrared attractive fixed points, where their RG evolution halts, the maximum of the distribution $P(g,t)$ can even \emph{cross the fixed-point value in its evolution} and is in general not monotonically attracted towards it during the RG flow. 
We have analytically exemplified this general result for the hypercharge coupling in the SM.

A second important application of our formalism relates to the propagation of errors under the RG flow. We have used the metastability scale in the SM Higgs sector as an example to demonstrate that individual RG trajectories do not contain sufficient information to propagate errors, because error propagation is intrinsically linked to the RG evolution of a distribution.
Our work thus provides the basis for a Bayesian assessment of (Beyond-) Standard-Model settings and UV completions.

\section*{Acknowledgments}
A.~E.~acknowledges the European Research Council's (ERC) support under the European Union’s Horizon 2020 research and innovation program Grant agreement No.~101170215 (ProbeQG).
This work has been funded by the Deutsche Forschungsgemeinschaft (DFG) under Grant No.~406116891 within the Research Training Group RTG 2522/1, to which A.~E.~is grateful for hosting her as a Mercator fellow during the conception of this work. 
A.~E.~has also been supported by a grant from Villum Fonden (29405). 

\bibliography{References}

\clearpage
\appendix

\begin{widetext}
\begin{center}
{\uppercase{\large{Supplementary Material}}}
\end{center}
\end{widetext}

\section{Derivation of the evolution equation for the probability distribution}

The evolution equation can be derived as follows: We consider an arbitrary function $u(g_i)$ of the couplings $g_i$, which has compact support, i.e., $u(g_i) \rightarrow 0$ for $g_i \rightarrow \pm \infty$. Its expectation value is given by
\be
\langle u(g_i) \rangle = \sum_i\int_{-\infty}^{+\infty}dg_i\, u(g_i)P(g_i,t).\label{eq:expone}
\ee
The time-derivative of this expectation value can be written in two ways. First, according to the definition of the expectation value
\begin{align}
\frac{d}{dt}\langle u(g_i) \rangle 
&= 
\left\langle \frac{d}{dt}u(g_i) \right\rangle= \left\langle \sum_i\frac{\partial u(g_i)}{\partial g_i} \beta_{g_i} \right\rangle
\nonumber\\
&=
\sum_i\int_{-\infty}^{+\infty}dg_i\ \frac{\partial u(g_i)}{\partial g} \beta_{g_i} P(g_i)
\nonumber\\
&= 
- \sum_i\int_{-\infty}^{+\infty}dg_i\ u(g_i) \frac{\partial}{\partial g_i}\left( \beta_{g_i} P(g_i)\right)
\;,
\end{align}
where in the last step we have used partial differentiation and the requirement that $u$ vanishes asymptotically. 

Second, by differentiating both sides of Eq.~\eqref{eq:expone} with respect to time,
\be
\frac{d}{dt}\langle u(g_i) \rangle =  \sum_i\int_{-\infty}^{+\infty}dg_i\, u(g_i) \frac{\partial P(g_i,t)}{\partial t}.
\ee

Setting both expressions equal and demanding that they hold for an arbitrary function $u(g)$ with compact support results in Eq.~\eqref{eq:partialtP}.
\\

\section{Formal solution via the method of characteristics}
\label{app:method-of-characteristics}

For completeness, we detail the solution of 
\begin{align}
	\frac{\partial P}{\partial t}= -\sum_i \left(
		\beta_{g_i} \frac{\partial P}{\partial g_i}
		+ \frac{\partial \beta_{g_i}}{\partial g_i}P
	\right)
	\;,
	\label{eq:PDE-reiterated}
\end{align}
provided some initial data 
$P(\vec{g}(t_0),t_0)=P_0(\vec{g}_0)$.

Applying the method of characteristics, we seek characteristic curves $\{t(s),\,\vec{g}(t(s))\}$, reparameterized by~$s$, along which the partial differential equation reduces to a set of coupled ordinary differential equations. Put differently, we seek $dg_i/ds=A_i(\vec{g},t)$ and $dt/ds=B(\vec{g},t)$ such that the reparameterized evolution equation of $P(s)=P(t(s))$ no longer contains partial derivatives with respect to $g_i(t(s))$ or $t(s)$, but is instead determined by a function $C(\vec{g},t)$ no longer containing derivatives, i.e.,
\begin{align}
	C \equiv \frac{d P}{ds} = 
	\sum_i \frac{dg_i}{ds}\frac{\partial P}{\partial g_i}
	+ \frac{dt}{ds}\frac{\partial P}{\partial t}
	\equiv
	\sum_i A_i\,\frac{\partial P}{\partial g_i}
	+ B\,\frac{\partial P}{\partial t}
	\;.
	\label{eq:total-differential}
\end{align} 
Comparing this condition with the original partial differential equation, we can identify
\begin{align}
	B(\vec{g},t) &\equiv 1 \equiv \frac{dt}{ds}\;,
	\label{eq:ode-parameterisation}
	\\
	A_i(\vec{g},t) &\equiv \beta_{g_i} \equiv \frac{dg_i}{ds}\;,
	\label{eq:ode-characteristics}
	\\
	C(\vec{g},t) &\equiv - \sum_i \frac{\partial \beta_{g_i}}{\partial g_i} P \equiv \frac{d P}{ds}\;.
	\label{eq:ode-solution}
\end{align}
Omitting an arbitrary integration constant, \cref{eq:ode-parameterisation} implies that $s=t$. As in the main text, we denote with $\vec{g}(t,\vec{g}_0)$, a formal solution of~\cref{eq:ode-characteristics} at time $t$, provided some initial conditions $\vec{g}_0$ at time $t_0$. Moreover, we assume that $\vec{g}(t,\vec{g}_0)$ is invertible for all $\vec{g}_0$ and denote the inverse by $\vec{g}_0(t,\vec{g})$.

With this inverse solution, we can obtain  the formal solution of~\cref{eq:ode-characteristics} by separation of variables, i.e.,
\begin{align}
	P(\vec{g},t) = 
	P_0\left(\vec{g}_0(t,\vec{g})\right)\cdot e^{\left[
		 - \int_{t_0}^{t} dt'\,{
		 	\sum_i \frac{\partial \beta_{g_i}}{\partial g_i}  
		 }
	\right]}
    \;.
	\label{eq:formal-solution-reiterated}
\end{align}
Here, we have integrated from $t_0$ to $t$ and identified the integration constant such that the solution realizes the initial data specified below~\cref{eq:PDE-reiterated}.
Whenever $\vec{g}(t,\vec{g}_0)$ solves \cref{eq:ode-characteristics}, invertibility is guaranteed, and $P(\vec{g},\,t)$ solves~\cref{eq:formal-solution-reiterated}, the latter also solves~\cref{eq:PDE-reiterated}, provided initial data $P_0(\vec{g}_0)$ at $t_0$.

\section{Initial distribution at the Planck scale}
\begin{figure*}[!t]
\begin{centering}
\includegraphics[width=\linewidth]{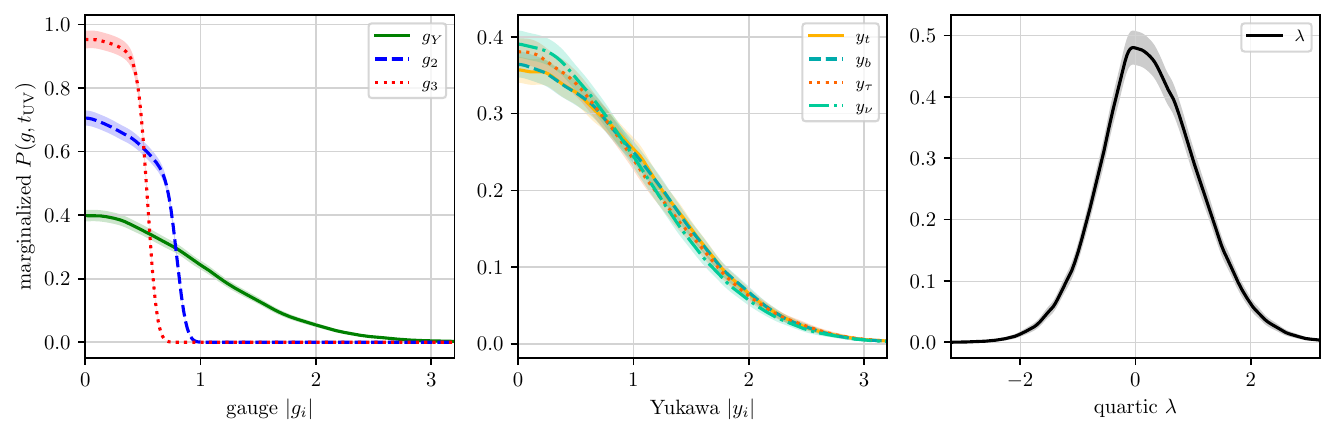}
\end{centering}
\caption{
\label{fig:SMnumericalflow-UV}
We show the initial UV probability distribution (upper row) and the final IR distribution (lower row), both projected/marginalized down to their dependence on individual couplings. The UV sample is drawn from a multidimensional Gaussian distribution with unit width for all couplings from which we discard all cases which lead to divergences in the RG flow (associated to IR Landau poles) between the UV and the IR scale, see main text for further explanation. For comparison, we also highlight the SM values (see vertical lines in the lower panel).
}
\end{figure*}

Here, we provide details on the initial distribution for the 
third-generation Standard Model evolved from the Planck scale $k_0=k_\text{Planck}=10^{19}\,\text{GeV}$ down to the electroweak scale $k=k_\text{ew}=173\,\text{GeV}$.

As remarked in~\cref{app:method-of-characteristics}, the validity of the formal solution in~\cref{eq:formal-solution-reiterated} relies on invertibility of all individual RG trajectories solving~\cref{eq:ode-characteristics}. This requires us to exclude trajectories for which the RG evolution diverges at intermediate scales. 
In case of the SM (evolved at one-loop and from the UV to the IR), such divergences (IR Landau poles) occur for the non-Abelian gauge couplings and for Higgs-quartic coupling. 

In practise, we start by sampling a Gaussian distribution with vanishing mean and unit variance in all coupling directions, i.e., an isotropic, multi-dimensional Gaussian distribution. Whenever a divergence occurs, we remove the corresponding initial coupling values from the initial UV distribution. As a result, our actual initial UV distribution does not remain Gaussian, see~\cref{fig:SMnumericalflow-UV}, but drops sharply to zero at larger values of the non-Abelian gauge couplings and becomes slightly skewed for the Higgs quartic coupling. Nevertheless, it remains broad and its only maximum remains at zero.
\\

\vfill

\section{Difference between quarks and leptons}
\begin{figure}[!t]
\begin{centering}
\includegraphics[width=0.8\linewidth]{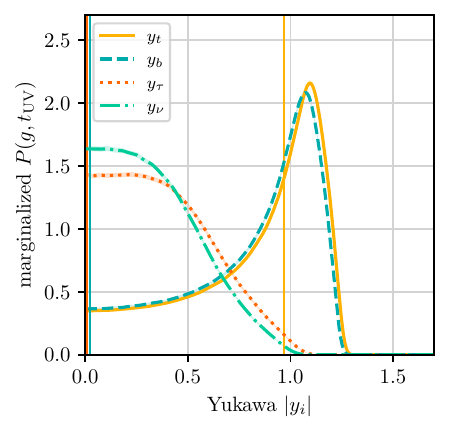}
\end{centering}
\caption{
\label{fig:SMnumericalflowprior}
As in the lower middle panel of~\cref{fig:SMnumericalflow}, but for fixed gauge evolution, corresponding to trajectories that realize the observed values at the electroweak scale.
}
\end{figure}
The underlying mechanism that separates the distributions of quark and lepton Yukawa couplings is that the SU(3) gauge coupling only couples to the quarks, and increases their Yukawa couplings, whereas the U(1) hypercharge coupling distinguishes between the four different fermions. Thus, we expect that the qualitative feature of a separation between quark Yukawas and lepton Yukawas is more pronounced at large values of SU(3). To exhibit this, we work with a strong theoretical prior and choose delta distributions for the initial condition for the three gauge couplings. As a result, there is a clearer separation between the distributions for quarks and for leptons in the IR, see Fig.~\ref{fig:SMnumericalflowprior}.

\end{document}